\documentclass[12pt]{article}
\usepackage{amsmath}
\usepackage{graphicx,psfrag,epsf}
\usepackage{enumerate}
\usepackage{url} 
\usepackage{amsfonts}
\usepackage{amssymb}
\usepackage{amsthm}
\usepackage{newlfont}

\theoremstyle{remark}

\usepackage{rotating}
\usepackage{longtable}
\usepackage{pdflscape}
\usepackage{tabu}
\usepackage{blindtext}
\usepackage{ifpdf}
\usepackage{tabularx}
\usepackage{graphicx}
\usepackage{epstopdf}
\usepackage{breqn}
\usepackage[superscript]{cite}
\usepackage{setspace}
\usepackage{ulem}

\usepackage{caption}
\usepackage{rotating}
\usepackage{booktabs,makecell}
\usepackage{multirow}
\usepackage[table]{xcolor}
\usepackage{hyperref}
\usepackage{amsmath}
\usepackage{cleveref}

\usepackage{dcolumn}
\usepackage{tikz}
\usetikzlibrary{shapes}
\usepackage{changepage}
\usepackage{epstopdf}

\newcolumntype{+}{!{\vrule width 1pt}}

\usepackage{array}

\date{}

\begin{document}

\def\spacingset#1{\renewcommand{\baselinestretch}%
{#1}\small\normalsize} \spacingset{1}

 
 \title{\bf Critical community size for COVID-19 - a model based approach to provide a rationale behind the lockdown}
\maketitle

\author{Sarmistha Das\textsuperscript{1} PhD, Pramit Ghosh\textsuperscript{2} MD, Bandana Sen\textsuperscript{3} PhD, and Indranil Mukhopadhyay\textsuperscript{1,*} PhD}
\bigskip
\\
\textsuperscript{1} Human Genetics Unit, Indian Statistical Institute, Kolkata, West Bengal, India
\\
\textsuperscript{2} Purulia Medical College, Purulia, West Bengal, India
\\
\textsuperscript{3} All India Institute of Hygiene \& Public Health, Kolkata, West Bengal, India
\\
* Corresponding author email: indranil@isical.ac.in; telephone: +91 9433 901325 
\bigskip

\section*{Summary}

{\bf Background} Restrictive mass quarantine or lockdown has been implemented as the most important controlling measure to fight against COVID-19. Many countries have enforced 2 - 4 weeks' lockdown and are extending the period depending on their current disease scenario. Most probably the 14-day period of estimated communicability of COVID-19 prompted such decision. But the idea that, if the susceptible population drops below certain threshold, the infection would naturally die out in small communities after a fixed time (following the outbreak), unless the disease is reintroduced from outside, was proposed by Bartlett in 1957. This threshold was termed as Critical Community Size (CCS). \\
{\bf Methods} We propose an SEIR model that explains COVID-19 disease dynamics. Using our model, we have calculated country-specific expected time to extinction (TTE) and CCS that would essentially determine the ideal number of lockdown days required and size of quarantined population.\\
{\bf Findings} With the given country-wise rates of death, recovery and other parameters, we have identified that, if at a place the total number of susceptible population drops below CCS, infection will cease to exist after a period of TTE days, unless it is introduced from outside. But the disease will almost die out much sooner. We have calculated the country-specific estimate of the ideal number of lockdown days.\\
{\bf Interpretation} Thus, smaller lockdown phase is sufficient to contain COVID-19. On a cautionary note, our model indicates another rise in infection almost a year later but on a lesser magnitude. \\
{\bf Funding} No funding to disclose.\\

{\it Keywords:} Critical community size, COVID-19, lockdown, quarantine, SEIR model

\section*{Introduction}

{\bf Research in context}
Several countries have resorted to restrictive mass quarantine or lockdown as it has been proved to be the most effective strategy to fight against the prevailing SARS-CoV2 or COVID-19 pandemic. To prevent stage III spreading of the virus or human-to-human transmission, various countries have imposed variable number of lockdown days that ranges from $2-4$ weeks. Depending on the real-time situation, various nations are updating on their lockdown period. However, it was not possible to determine or predict accurately the ideal number of lockdown days in advance as we are all fighting against a novel virus. Still, based on the nature of pandemics and epidemics in the past and the successful mode of encounter with them, the human race has learnt how to fight out the disease, in absence of any specific treatment. Mostly on the basis of the average incubation period of SARS-CoV2 virus, countries have devised the phase of lockdown. In the present day, the COVID-19 pandemic figures from developed countries have surpassed the figures from China, from where this infection had originated. In such a dire situation, it is very difficult to propose quintessential lockdown period specific to any country. The whole world is struggling to obtain unbiased data to predict on the pandemic. But even then many questions arise in our minds as ``Will the implemented number of lockdown days eradicate the virus?" or ``Will it come soon again after the lockdown is over?" or like ``How soon will the next wave of infection occur, if it comes back again?" Answer to these questions lies in the idea of critical community size (CCS) from Soper \cite{soper1929interpretation} and Bartlett's \cite{bartlett1957measles, bartlett1960critical} works in 1920s and 1950s. \\

{\bf Added value of this study} 
Our study provides a rationale behind the determination of the lockdown period in different countries going through the catastrophic effect of the pandemic. Moreover, we provide country-wise expected time to extinction (TTE) of the disease and the number of people who could safely stay together or quarantine themselves together in clusters. The most interesting observation from our study is that the lockdown of 2 weeks or more (country-specific) will eradicate the disease almost completely. We term this period as Temporary Eradication of Spread Time (TEST) for disease, which is much less than the TTE of the disease. But there will be more waves of infection after different period of time in different countries. Soothing part from our deduction is that the virus will not be able to create any havoc on its return. \\

{\bf Implications of all the available evidence}
In this scenario of utter dilemma, with the available world-wide data, we have provided country-wise estimates of the ideal lockdown phases using our proposed mathematical model. To the best of our knowledge, any guideline for country-wise mathematical prediction of lockdown days is not available till date. So, as the famous British statistician George E. P. Box pointed out, ``All models are wrong but some are useful", we only hope that our deductions will provide some helpful suggestion to the world, where we are all struck in the pandemic. \\

In the face of COVID-19 pandemic, many countries have implemented restrictive mass quarantines or lockdown as the primary controlling measure to confine the number of secondary transmission of the disease within countries. In absence of any specific medicine to treat the disease, patients are given supportive care to help them breathe. But given the mode of transmission of the epidemic, health care systems of even developed countries are crumbling down within a week or two. So along with many other countries densely populated India has resorted to complete lockdown for $3$ weeks to prevent stage III transmission of the disease. Available data confirms that the pandemic has affected a million people in around $200$ countries till date and already claimed more than $0.05$ million lives across the world within approximately two months. After World Health Organisation (WHO) declared the outbreak as a pandemic, many countries initiated partial to complete lockdown as done in some provinces of China after the outbreak started and by the end of March, one-third of the global population is under some form of lockdown. 

Many countries implemented variable number of lockdown days, but no country has come up with any magic figure for the ideal period of lockdown. No clear-cut guideline or rationale behind the number of lockdown days has been announced by any country or WHO till date to the best our knowledge. The period of $2$ - $4$ weeks is determined mostly on the trial and error basis. The prediction on the number of trial lockdown days could be possibly and partially based on the fact that, an affected individual could be contagious in the first $14$ days of contracting the disease and the number of known positive cases at the time of taking decision.

Idea of quarantining a small group of people after an epidemic outbreak to arrest the disease dates back to 1950s when English statistician M.S. Bartlett introduced critical community size. Bartlett \cite{bartlett1957measles, bartlett1960critical} proposed the idea that if the susceptible population is below a certain some threshold, the infection is as likely as not to die out after a period of time (after the epidemic outbreak) in small communities, unless the disease is reintroduced from outside. Bartlett termed this threshold as Critical Community Size (CCS). In the present context, CCS could guide government / health policy makers with an objective strategy of lockdown period as opposed to subjective trial and error phases of lockdown. 

After an epidemic outbreak in a community, the infection persists long enough to engulf the entire susceptible population. Local extinction of the disease could be possible if the susceptible population gets depleted. In large communities, the tendency of eventual damp down of the recurrent epidemics is balanced by random variability. But in small communities the infection would die out when the number of susceptible falls below a certain threshold, which is the CCS. Only a limited number of works\cite{naasell2005new, andersson2000stochastic} including our work (under review) are available on CCS, may be because it involves complicated calculations even for simplest mathematical model viz. SI (S: Susceptible, I: Infected) model. However, since the actual extent of an epidemic can be assessed only retrospectively, it is essential to calculate the CCS for COVID-19 based on a realistic model that depends on the parameters which could be determined for a specific locality. 

We propose an SEIR (S: Susceptible, E: Exposed, I: Infected, R: Recovered) model to explain the disease dynamics of COVID-19. We have derived with evidence the rationale behind the importance and extent of the lockdown period and also the number of people who could safely stay together in this lockdown phase. In absence of much prior knowledge on the disease, we have to rely on the mathematical predictions to combat the virus. In this article, we provide a cautionary note from the mathematical deductions, that this pandemic will last more than an year and may start afresh again. Our work resonates the leaked news in the American daily The New York Times (https://www.nytimes.com/2020/03/17/us/politics/trump-coronavirus-plan.html) that official authority as well as experts anticipates that the pandemic ``will last 18 months or longer" and could include ``multiple waves" of infection. 

But there is always a ray of hope. This nature of epidemic is also seen previously after SARS or MERS outbreak. Our deduction reveals that there is nothing to panic as the lockdown, if properly followed, could almost wipe out the disease and the next wave of infection will not really recur as an alarming pandemic. Although we have to bear the burden of economic sluggishness or downfall, the COVID-19 epidemic could be controlled and hopefully won't create a public health emergency in the near future.

\section*{Methods}
We propose an SEIR model to explain the dynamics of COVID-19 infection. The entire population is divided into four compartments. These compartments are mutually exclusive in the sense that no person can belong to more than one compartment at any time point. The four compartments are: susceptible individuals (S), individuals with and without symptoms of the disease but not yet tested positive for COVID-19 (E), infected individuals who are clinically tested positive (I), and individuals who are known to have recovered from the disease (R). Note that an individual belonging to class E may transmit the disease during the incubation period, which is approximately 1-14 days. Under this situation, we consider the model as:

\begin{eqnarray}
\frac{dS}{dt} & = & \Lambda - \beta\left(I(t) + \phi E(t) \right) \frac{S(t)}{N} - \mu S(t)\\
\frac{dE}{dt} & = & \beta\left(I(t) + \phi E(t) \right) \frac{S(t)}{N} - (\gamma + \mu)E(t)\\
\frac{dI}{dt} & = & \gamma E(t) - (\delta + \mu +d) I(t)\\
\frac{dR}{dt} & = & \delta I(t) - \mu R(t)
\end{eqnarray}

Here $\beta$ represents the contact rate for COVID-19 transmission from infected to susceptible individuals, $\phi$ is the proportion of the exposed who could transmit the disease, an individual in E moves to I at the rate $\gamma$, $\delta$ is the recovery rate, $d$ is death rate due to the disease and $\mu$ is the natural birth/death rate in the population. Here, $\Lambda=\mu N(t)$ where $N(t)$ is the population size at time $t$.

The most important quantity in this type of model is the basic reproduction number, which is:
\begin{align}
R_0 = \frac{\beta (\phi (\delta + \mu + d) + \gamma)}{(\gamma + \mu) (\delta + \mu + d)}
\end{align}

Our method of evaluation of CCS depends on the parameters  of the model and it involves a series of mathematical derivation and processes (See Supplementary Material). First we develop a fully stochastic model corresponding to the deterministic model (1)-(4). Assuming quasi-stationarity and non-extinction of infection, we derive expected TTE {\it i.e }$E(\tau_Q)$ of the disease. $E(\tau_Q)$ involves some probability terms that we evaluate using diffusion approximation of the scaled state variables (S, E, I, R). $\tau_Q$ is a function of the CCS. We derive quasi-period $\hat{T}_0$ (say) in terms of angular frequency that is obtained using linearised system at equilibrium points. Then using the relation $E(\tau_Q)\log 2 = \hat{T_0}$\cite{naasell2005new}, we could finally evaluate the CCS for the system. Note that this CCS is a function of the parameters involved in the model that vary from country to country, even from one region to another. Thus CCS is not any unique value for all regions or countries.

Note that the calculation of CCS involves a series of complicated processes and mathematics that demand few approximations at some stages. So we also did simulations with realistic parameter values for different countries. The simulation study matches with the theoretical findings.

\section*{Results}

For COVID-19 transmission, we have calculated the CCS and TTE of the disease based on our proposed SEIR model. Note that the value of the CCS is approximate as we have made some mathematical approximation while applying diffusion approximation to find the quasi-stationary distribution. The value of CCS for a community or a country, depends on the model parameters specific to that community. We deduce these parameters from the available information on COVID-19 till date.

Actual fatality rate due to any epidemic could only be calculated after the epidemic gets over. But in the middle of the pandemic, it is difficult to assess. So we determine the country-specific death rate (d) based on the number of deaths up to April 01, 2020 and the total number of infected individuals till $7$ days prior to this date. Here we assume that COVID-19 patients do not die within the first seven $7$ days after detection. 

Different countries have implemented varying criteria of discharging COVID-19 patients from hospitals (\url{https://www.ecdc.europa.eu/sites/default/files/documents/COVID-19-Discharge-criteria.pdf}) making the actual recovery rate very tricky to calculate amid the pandemic. It is yet unknown whether all the discharged patients have fully recovered from the disease or they got sick again shortly after (\url{https://www.cdc.gov/coronavirus/2019-ncov/hcp/faq.html}). So we have assumed the recovery rate ($\delta$) based on the number of recovered patients up to April 01, 2020 and the total number of infected individuals till $14$ days prior to this date. The country-specific death and recovery rates are obtained from \url{https://www.worldometers.info/coronavirus/}. 

On exposure to COVID-19 infected patient when a person gets infected, the virus itself makes enough copies of itself in the host (that is the person who was exposed to the infected patient) within some time, which the host begins to shed through coughs or sneezes or other transmission methods. Assuming that within the time taken by the virus to replicate itself, the host will not shed the virus out of its body and hence the host will not be contagious within this period. This time is usually little more than one day of the exposure. So, we have assumed that after day 1 of the exposure to the virus, the host becomes contagious, {\it i.e} an exposed person has $93\%$ chance of transmitting the infection as that of an infected person. Since the incubation period of COVID-19 is $14$ days, we take $\phi=1-\frac{1}{14}=0.93$. 

Another very tricky and country dependent parameter is the rate of detection of positive cases from among the exposed pool of people, {\it i.e.} percentage of exposed people that are actually tested to be COVID-19 positive. In absence of manpower and testing kits in such a dire situation, we will not be able to know the actual proportion of the exposed who could later on become COVID-19 patient. We have assumed it to be $30\%$ (across the world) that is, $\gamma=0.3$. The rate of natural birth or death is taken as $\mu=0.015$ that stabilises the population under normal scenario. In absence of actual contact rate ($\beta$), we have assumed $\beta$ over a feasible range of values.

We represent the country-wise CCS and expected TTE of the disease in Figure $1$. For example, the CCS for Belgium (represented by the brown line in Figure 1A) lies between $34$ and $52$ while the expected TTE of the disease (represented by the brown line in Figure 1C) varies between $500$ and $693$ days for different contact rates in the range $0.31$ to $0.37$. Therefore, for a particular contact rate $\beta=0.37$ say, the CCS of Belgium is $34$ means that, if the susceptible population of any community in Belgium is below $34$, the infection will die out unless it is re-introduced from outside after the corresponding expected TTE, which is $599$ days.

\begin{figure}[h!]
\begin{center}
\includegraphics[width=4.7in]{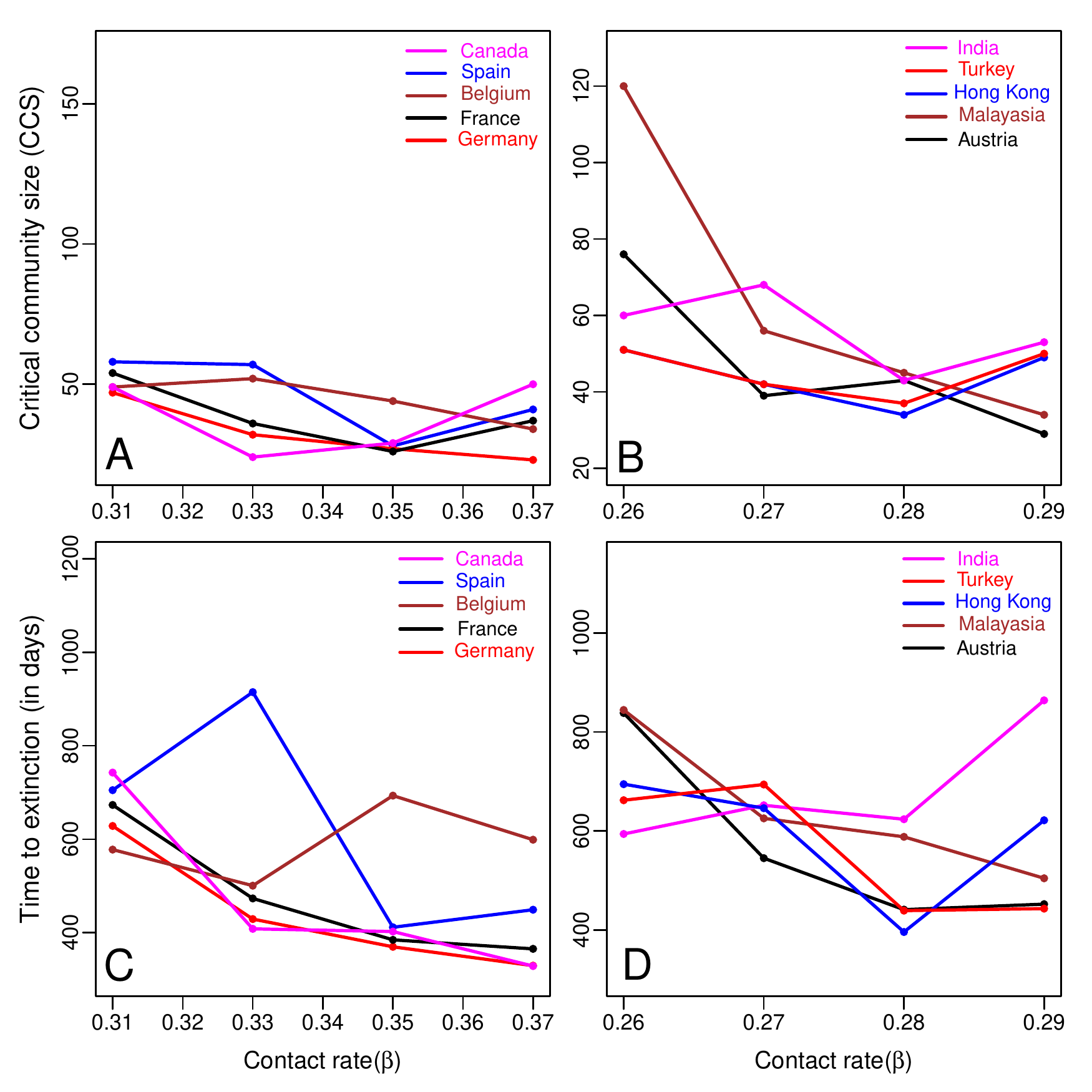}
\end{center}
\caption*{\footnotesize{Figure 1: (A, B) provide country-specific CCS values for a range of contact rates ($\beta$). (C, D) give country-specific expected TTE values for a range of contact rates ($\beta$). The countries are color coded as given in the legend in each plot.}\label{fig:first}}
\end{figure}

But we observe that the number of infections falls below $0.5$ after around $16$ days when the community size of quarantined people drops to the CCS value $34$ (as shown in Figure 2). Thus in this case the disease gets nearly eradicated after $16$ days, but it may come back again like any other epidemic disease before actually becoming extinct. However, after TEST it will not likely to create a pandemic as in the current situation. We observe that the number of infections drops below $0.5$ after around $15-30$ days in most of the countries. For example, in India we observe that TEST is $24$ days. (Figure 3). So even if the expected TTE of the disease is more than an year, it will almost cease to exist after a few days if the rates of transmission of the disease could be kept low as in the prevailing lockdown situation. 

The CCS and expected TTE for a few other countries like Spain, Germany, France and Canada are given in Figure 1A \& 1C, along with Belgium respectively. Figure 1B \& 1D represent respectively the CCS and expected TTE of countries viz. India, Austria, Malaysia, Hong Kong and Turkey.

\begin{figure}[h!]
\begin{center}
\includegraphics[width=4.7in]{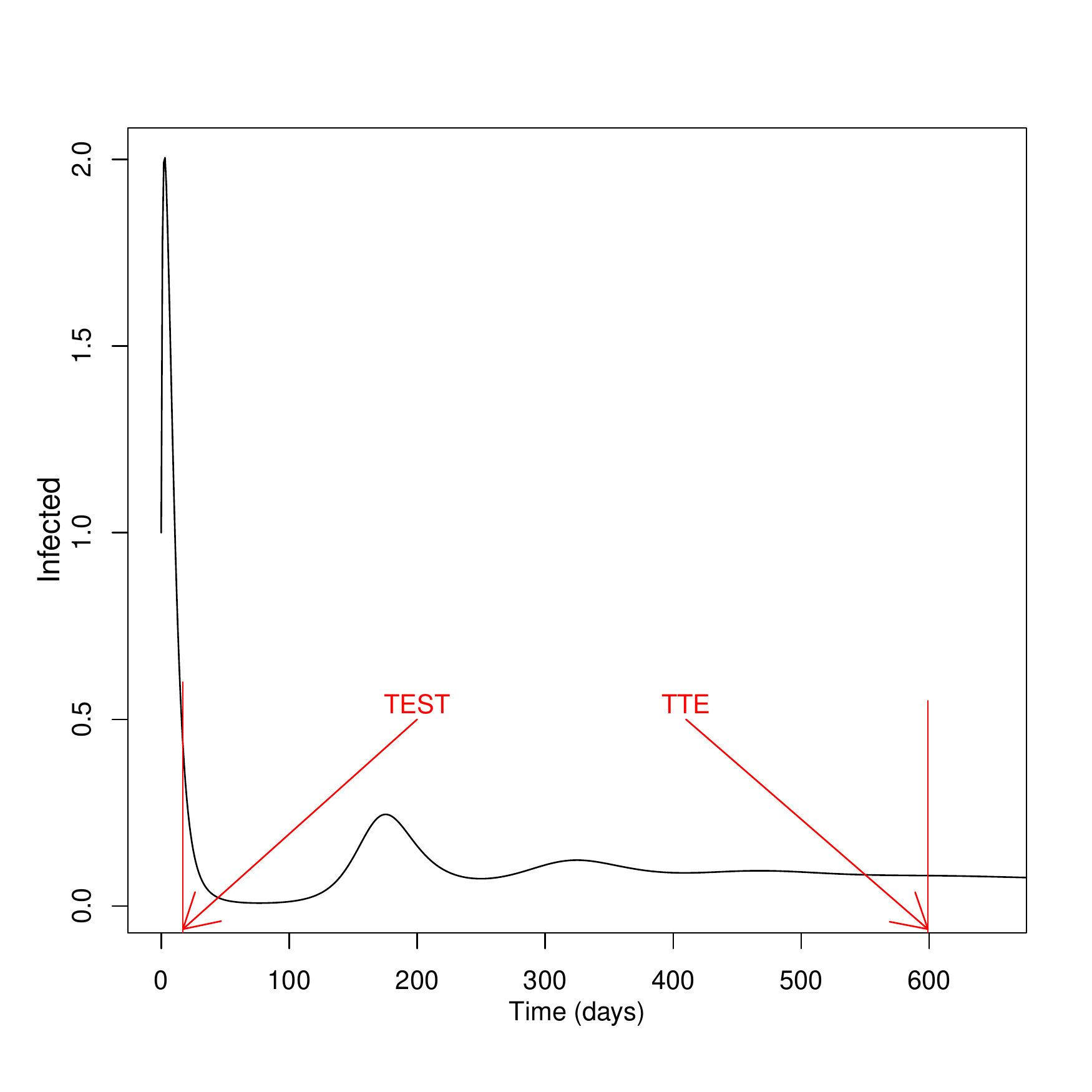}
\end{center}
\caption*{\footnotesize{Figure 2: TTE of Belgium is 599 days corresponding to CCS of 34 but the disease almost becomes extinct at TEST that is around 17 days. Multiple waves of infection are likely to occur before TTE but that will be unable to create a pandemic as in the current situation.}\label{fig:second}}
\end{figure}

\begin{figure}[h!]
\begin{center}
\includegraphics[width=4.7in]{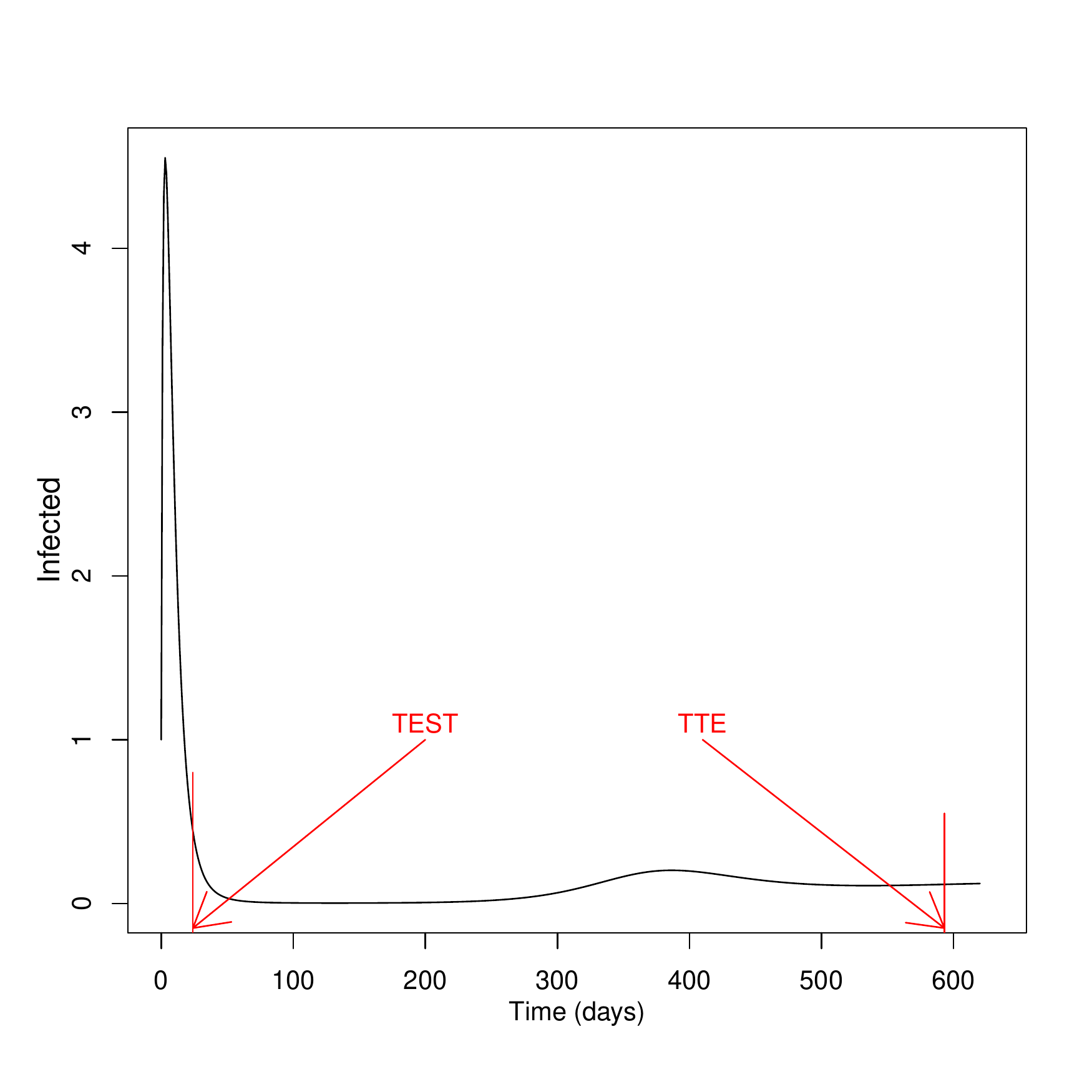}
\end{center}
\caption*{\footnotesize{Figure 3: For India, the TEST or required lockdown period is 24 days but the estimated TTE is around 593 days. India is likely to have another wave of infection after around a year but that will be unable to create a pandemic as in the current situation.}\label{fig:third}}
\end{figure}

We present country-specific objective lockdown periods based on our model in Table 1.

\begin{table}[!ht]
\begin{adjustwidth}{-0.25in}{0in} 
\centering
\caption{
{\bf Chart of ideal lockdown period (in days) required for the countries}}
\begin{tabular}{|c|lc|lc|c}
\hline
{\text{\bf Country}} &{\bf min.TEST} &{ \bf max.TEST} &{ \bf mean.TEST} &{ \bf sd.TEST}\\
\hline
\text{Italy} & 24	 & 32	 & 29 & 3.46\\
\hline
\text{Spain} & 15 & 22 & 18.83 & 2.17\\
\hline
\text{USA} & 14 & 25 & 19.45 & 3.08\\
\hline
\text{Germany} & 15 	& 25	 & 19.82 & 3.25\\
\hline
\text{France} & 16  & 26 & 19.67 & 2.96\\
\hline
\text{UK} & 49 & 53 & 51 & 2.83\\
\hline
\text{Switzerland} & 19 & 27 & 22.86 & 2.73\\
\hline
\text{Belgium} & 16 & 22 & 19.7 & 2.58\\
\hline
\text{Austria} & 19 & 28 & 23.33 & 3.33\\
\hline
\text{Denmark} & 16 & 25 & 20.8 & 2.9\\
\hline
\text{Malaysia} & 19 & 28 & 21.5 & 2.98\\
\hline
\text{Phillipines} & 30 & 37 & 34 & 3.61\\
\hline
\text{Poland} & 45 & 54 & 48.67 & 4.73\\
\hline
\text{India} & 22 & 29 & 25.14 & 2.41\\
\hline
\text{Indonesia} & 25 & 32 & 28.75 & 2.99\\
\hline
\text{Greece} & 43 & 57 & 48.33 & 7.57\\
\hline
\text{Algeria } & 19 & 27 & 23 & 2.67\\
\hline
\text{Hong Kong} &22 & 29 & 24.67 & 2.42\\
\hline
\text{Turkey} & 23 & 31 & 25.83 & 2.79\\
\hline
\text{Canada} & 15 & 21 & 18.5 & 2.27\\
\hline
\text{Portugal} & 44 & 59 & 49.33 & 8.39\\
\hline
\text{Brasil} & 43 & 47 & 44.33 & 2.31\\
\hline
\text{Israel} & 27 & 33 & 30.75 & 2.63\\
\hline
\text{Australia} & 30 & 33 & 31.5 & 2.12\\
\hline
\text{Sweden} & 51 & 64 & 56.33 & 6.81\\
\hline
\end{tabular}
\label{table1}
\end{adjustwidth}
\end{table}

\section*{Discussion}

The earlier the lockdown starts, the better it is for any community. With the delay in the number of days to initiate lockdown, time for the disease to slow down will increase in the absence of any specific treatment to COVID-19. Before any country enters stage III of the epidemic, in absence of vaccination or any specific treatment, complete quarantine will be able to contain the disease from spreading further. Our work suggests that if people are quarantined in small groups presented by country-specific CCS, the disease will become extinct after the corresponding expected TTE. But it becomes almost extinct after around TEST. So lockdown should to be taken very seriously to fight against the pandemic. This paper provides evidence of the fact that even after the lockdown phase, multiple waves of the disease will re-occur but it will not create pandemic situation. 

In the present situation where contact tracing of the infected has attracted world-wide attention to trace down the exposed individuals, we suggest to quarantine them in groups as small as country-specific CCS, so that after $15-30$ days (as specified for each country) the disease would become almost extinct unless any infection is re-introduced from outside. There could be some mathematical limitations due to some approximations in the calculation of CCS and expected TTE of the disease. But the approximations are inevitable and do not affect much the actual CCS or TTE as reflected from our numerical evaluation.

Another undeniable consequence of the current pandemic is the great negative impact on the economy. This is further magnified due to almost paralysed transactions in many sectors due to lockdown. Moreover, there is a fear among the general population that if any infection pops up, it might create another round of spread of the disease. This fear, which is not completely unrealistic, may extend the lockdown further. However, after the scheduled lockdown period, if any individual gets infected and a few others get exposed to that person, we need to check whether the total number of such individuals is less than the CCS. If so, this group needs to be quarantined and the rest of the population would be free from immediate danger. However, if the exposed group is larger than the CCS, we recommend to split them in subgroups, each of size at most CCS and quarantine them separately. However, careful and detail surveillance is required to trace all such exposed and infected individuals. Moreover, this localised lockdown or quarantine would help flowing the daily life and might prevent, at least to some extent, the economy from being further weak.

CCS has another implication in allocation and arrangement of hospital beds in the quarantine/hospital facilities. If the CCS is $30$ (say), a block of $30$ or less number of beds should arranged so that total movement within this block, including patients and healthcare people should not exceed that number. Otherwise if one big room accommodates a large number of patients, secondary infection may start spreading. Our model is capable of estimating the average number of testing kits and hospitalisations required in advance to some extent, depending on the available data.

However, lockdown in its truest sense may not be feasible in a vast and diverse country like India. So targeted or  focused lockdown of selected high risk population might be a cost effective option compared to a generalised lockdown. This would mean saturating screening for cases and identifying the critical number of contacts.

So the take away message in the middle of the COVID-19 pandemic is that till the end of expected TTE (country-specific) of the disease we have to be careful. With further onset of COVID-19 symptoms in the future, we should follow quarantine rules strictly. Like any other epidemic, COVID-19 has the tendency to recur but it will not create any alarming pandemic in the future provided we keep a vigilant eye on our hygiene and have vaccination and treatment in the future.

\section*{Conflict of interest}
The authors do not have any conflict of interest.

\bibliographystyle{vancouver}

\bibliography{Das_manuscript_COVID19.bib}

\end{document}